# Spin filtering effect in intrinsic 2D magnetic semiconductor $Cr_2Ge_2Te_6$

Honglei Feng[1,2], Gang Shi[1,2], Dayu Yan[1,2], Yong Li[1,2], Youguo Shi[1,2,3], Yang Xu[1,2], Peng Xiong[4], and Yongqing Li[1,2,3, a]

1. Beijing National Laboratory for Condensed Matter Physics, Institute of Physics, Chinese Academy of Sciences, Beijing 100190, China
2. School of Physical Sciences, University of Chinese Academy of Sciences, Beijing 100190, China
3. Songshan Lake Materials Laboratory, Dongguan, Guangdong 523808, China
4. Department of Physics, Florida State University, Tallahassee, FL 32306, USA

a) Author to whom correspondence should be addressed: yqli@iphy.ac.cn.

**Abstract:** All van der Waals (vdW) $Fe_3GeTe_2$/$Cr_2Ge_2Te_6$/graphite magnetic heterojunctions have been fabricated via mechanical exfoliation and stacking, and their magnetotransport properties are studied in detail. At low bias voltages large negative junction magnetoresistances have been observed and are attributed to spin-conserving tunneling transport across the insulating $Cr_2Ge_2Te_6$ layer. With increasing bias, a crossover to Fowler-Nordheim tunneling takes place. The negative sign of the tunneling magnetoresistance (TMR) suggests that the bottom of conduction band in $Cr_2Ge_2Te_6$ belongs to minority spins, opposite to the findings of some first-principles calculations. This work shows that the vdW heterostructures based on 2D magnetic insulators are a valuable platform to gain further insight into spin polarized tunneling transport, which is the basis for pursuing high performance spintronic devices and a large variety of quantum phenomena.

Recent advances in the study of 2D vdW magnetic systems have provided exciting opportunities in pursuing spintronic devices[1-4]. Among them the work on vdW magnetic insulators is of particular interest, since it may help to overcome some critical difficulties encountered in utilizing magnetic insulators for spin filtering and magnetic proximity effects. Spin polarized tunneling in Eu chalcogenides was first evidenced as spin-selective Fowler-Nordheim tunneling in 1960s[5]. Moodera *et al.* later demonstrated that these materials could serve as very efficient spin filters with Tedrow-Meservery type tunneling spectroscopy[6,7]. This was followed by a lot of efforts on investigating spin filtering effects in $NiFe_2O_4$[8,9], $CoFe_2O_4$[10,11] and other oxides[12,13], which have much higher Curie temperatures than the Eu chalcogenides and thus offer a potential route to developing practical TMR devices. However, the obtained TMR ratios at room temperature have been limited to only several percent. This has been attributed to the defect formation in the oxide layer due to lattice mismatch[14] as well as other issues.

2D vdW magnetic insulators possess atomically flat surfaces and highly uniform thicknesses, and interact relatively weakly with other 2D materials when assembled into heterostructures[1-4]. These advantages make the 2D magnetic insulators very appealing for realizing ideal spin filtering devices. Impressive progress has been made on electron tunneling across thin layers of $CrI_3$ and $CrBr_3$ in the heterojunctions with *nonmagnetic* electrodes[15-17]. Ferromagnetic ordering temperatures in these materials are, however, much lower than room temperature. Large scale calculations based on the first principles have been performed to predict high-$T_C$ 2D magnetic insulators[18,19]. Unfortunately, the effectiveness of such computational effort is still hampered by complicated exchange-correlation interactions in the vdW magnetic insulators[20].

For example, $Cr_2Ge_2Te_6$ (CGT) is an intrinsic magnetic semiconductor with perpendicular magnetic anisotropy and possessing excellent insulating properties at low temperatures, and has become a very attractive platform for investigating various magnetic proximity effects[21,22]. Despite lots of research, there is still no consensus on two key aspects of the electronic structure of CGT. One is the size of the energy gap: reported experimental values[23-25] spread from 0.19 eV to 0.74 eV, and similar disparity is also present in the computed values[20]. The other is how the conduction band is spin-split: while several groups[23,26,27] found the conduction band minimum residing in a minority spin band, others[24,25] claimed it belonging to majority spins. No experiment has been reported so far on determination of the sign of conduction band spin-split in CGT. The insufficient knowledge on the band structure prevents a definitive understanding of spin polarized tunneling across the magnetic insulator.

In this paper, we report on an experimental study of the magnetic tunnel junctions constructed with microflakes of $Fe_3GeTe_2$ (FGT), CGT, and graphite, in which FGT has perpendicular magnetic anisotropy and $T_C$ of about 215 K[28,29] and serves as a ferromagnetic metal electrode, CGT is a spin filter, and graphite works as a nonmagnetic counter-electrode. Negative junction magnetoresistances with magnitude up to at least 30% have been observed. Temperature and bias-voltage dependences of the transport properties coherently point to a spin-polarized tunneling process, in which the spin-split conduction band of the CGT layer is responsible for the negative spin filtering effect.

The spin filtering device investigated in this work is schematically depicted in Fig. 1a. It consists of five microflake layers exfoliated from vdW materials, stacked in a sequence of $h$-BN, FGT, CGT, graphite, and $h$-BN, with the hexagonal boron nitride ($h$-BN) flakes working as encapsulation layers. Fig. 1b shows an optical image of device A, in which the CGT layer is 2.1 nm thick (equivalent to 3 CGT atomic sheets[30]), as well as the measurement scheme for the junction resistance $R_J$ and differential conductance ($G \equiv dI/dV$). The data shown below were collected from this device, unless specified otherwise. The data taken from other devices, as well as additional data from device A, are given in the supplementary material.

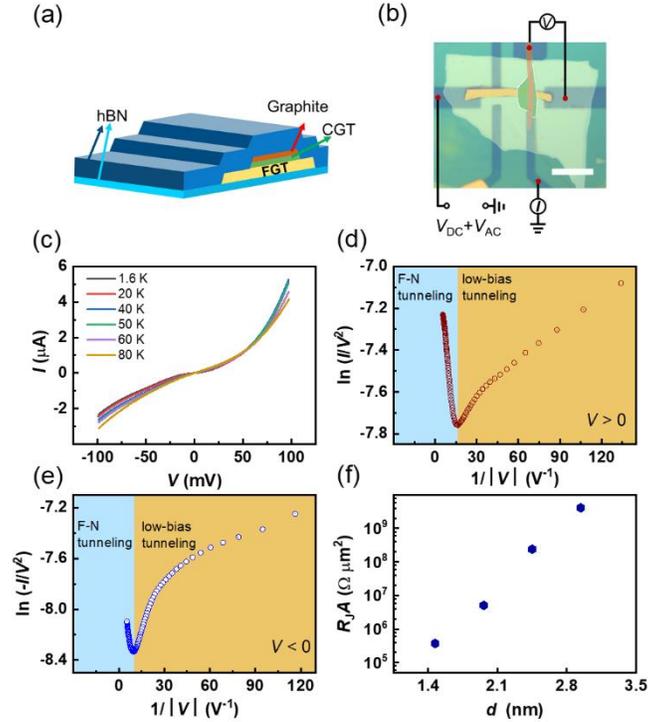

**Figure 1.** Basic characterization of the FGT/CGT/graphite heterojunctions. (a) Schematic illustration of the device geometry. (b) Optical image of a typical device (device A, in false color, scale bar: 10 μm). The boundary of the tunneling barrier (CGT flake) is outlined in white color. (c) *I-V* characteristics at $T = 0.3$ K for parallel spin alignment between the FGT and CGT layers. (d,e) $\ln(|I|/V^2)$ vs. $1/|V|$ for the positive (d) and negative dc biases (e), derived from the *I-V* curve shown in panel (c). (f) Thickness dependence of the product of junction resistance $R_J$ and junction area A at $T = 1.6$ K.

Both FGT[31,32] and graphite have very low resistivities and CGT becomes insulating at low temperatures[23]. The resistance of FGT/CGT/graphite junction is hence mainly determined by the transport process across the CGT layer. Depicted in Fig. 1c is an *I-V* curve of the heterojunction recorded at $T = 0.3$ K. The same set of data replotted in the form of $I/V^2$ vs. $1/V$ (Figs. 1d & 1e), is consistent with the behavior of a tunnel junction — crossover from the so-called direct tunneling at low bias voltages to the Fowler-Nordheim tunneling at high biases ($V > 70$ mV or $V < -103$ mV)[33,34]. Fig. 1f further shows that the junction resistance increases very rapidly as the thickness of the CGT layer, denoted as $d$, becomes larger. The data collected from four devices with $d = 2.1$ - 4.2 nm, corresponding to 3-6 atomic layers of CGT, follow $R_J A \propto \exp(\kappa d)$, where $A$ is the junction area, and $\kappa$ is a constant. Such an exponential dependence is also consistent with the tunneling transport, in which $\kappa$ is related to the decay rate of electron wavefunction in the tunnel barrier[35].

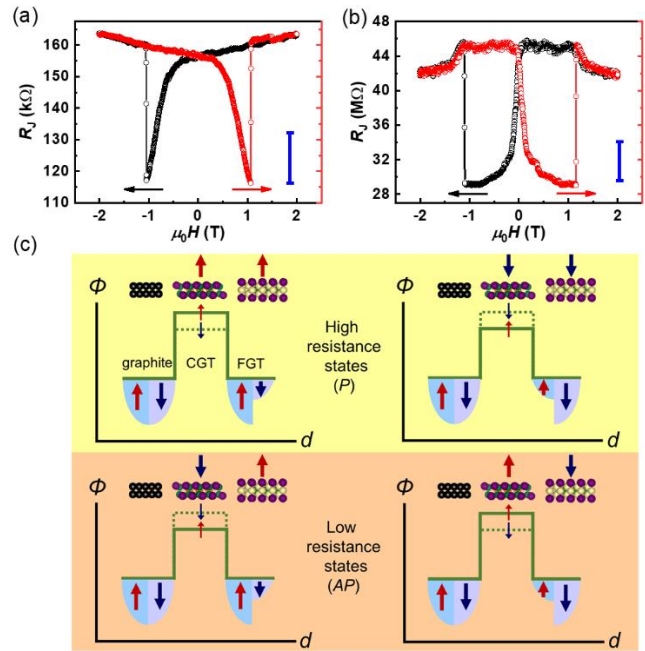

**Figure 2.** *I-V* characteristics and differential conductance spectra of the FGT/CGT/graphite heterojunction at various temperatures. (a) *I-V* curves at $T = 1.6$-80 K for parallel magnetization direction between FGT and CGT. (b) Dependence of differential conductance $G$ on the dc bias voltage $V$. The insets are the band diagrams for negative (left) and positive (right) biases. (c) Temperature dependence of $G$ at high bias voltages. The insets depict band diagrams of the $Cr_2Ge_2Te_6$ layer in the ferromagnetic (lower-left) and paramagnetic (upper-right) states. (d) Zero-bias conductance $G_0$ vs. temperature. The inset shows the band diagram for zero dc bias. An ac excitation of 1 mV was applied during the differential conductance measurements.

Fig. 2a shows that for the temperatures below 40 K, the *I-V* curves nearly collapse onto a single curve at bias voltage $V >70$ mV. As depicted in Fig. 2b, similar behavior also appears in the differential conductance spectra. Fig. 2c depicts that differential conductances at high biases ($V =85, 90, 95$ mV) vary very little at $T < 40$ K, but *decrease* significantly as the temperature further increases. This strongly suggests that thermally activated process should not play an important role in the transport across the CGT layer. Therefore, the temperature dependence observed at high biases further supports that the transport enters the Fowler-Nordheim tunneling regime, in which the conductance is determined by the profile of tunnel barrier[5,33,34]. The weak *T*-dependence at $T < 40$ K can be attributed to little change in the band structure of CGT, whereas at higher temperatures the reduced band splitting in CGT increases the height of tunnel barrier, leading to smaller *G* values, as illustrated in the insets of Fig. 2c.

In contrast to the weak *T*-dependence of the high-bias differential conductance at $T < 40$ K, the low-bias *G* varies strongly with the temperature. Fig. 2d shows that the zero-bias conductance, $G_0$, increases by ~140% as *T* increases from 1.6 K to 40 K. This can be ascribed to large mismatch between the Fermi wavevectors of graphite and FGT. Consequently, elastic tunneling cannot take place without compensation of the difference in momentum. At finite temperatures the tunneling process can be enabled by absorbing a magnon and simultaneously emitting another one that has the same energy but a different wavevector. Such a two-magnon-assisted tunneling mechanism has been invoked to explain the strong *T*-dependence of $G_0$ observed in a graphite/CrBr$_3$/graphite heterojunction[15]. Since the two magnons have the same angular momentum in this process, the spin is conserved, providing a route for spin polarized tunneling.

Fig. 3a depicts the junction resistance at $T = 1.6$ K as a function of magnetic field, which was applied in the out-of-plane direction throughout in this work. Starting from $\mu_0 H =+2$ T, which is higher than magnetization saturation fields of both FGT[36,37] and CGT[38], $R_J$ begins to drop slowly as the magnetic field decreases. When $H$ becomes negative, $R_J$ decreases at accelerating rates, until the coercive field of FGT ($\mu_0 H_c \approx -1.1$ T) is reached, which is signaled by an upward jump in $R_J$. For the up-sweeping magnetic field, there is a similar sharp change in $R_J$ due to the magnetization reversal in the FGT layer. No sharp jumps related to the magnetization switching in the CGT layer can be identified. This can be attributed to the soft magnetism in CGT, which has nearly zero coercive fields in both bulk[23] and 2D[38] forms.

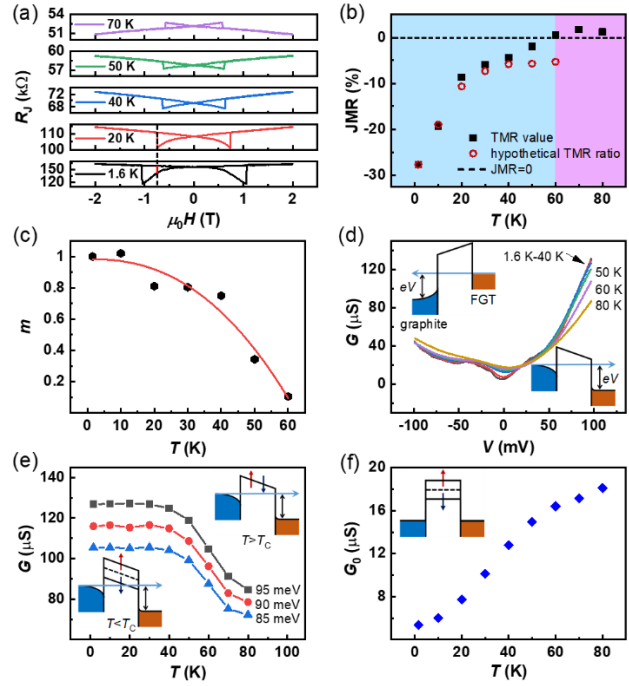

**Figure 3** Temperature dependence of TMR in the FGT/CGT/graphite heterojunction. (a) Magnetic field dependence of junction resistance. The horizonal arrows indicate the directions of magnetic field sweeps, and the blue scale bar represents a TMR value of 10%. The data were taken with zero dc bias and an ac excitation of 1 mV at $T = 1.6$ K. The middle and bottom panels are schematic diagrams for parallel and antiparallel magnetization configurations in the FGT and CGT layers. (b) Hysteresis curves of the junction resistance at selected temperatures. (c) $T$-dependence of the measured TMR (solid squares), and hypothetical TMR (open circles). (d) $T$-dependence of the ratio between the measured and hypothetical TMR values (solid hexagons) and the fit to a Bloch-law-like equation (line). An ac excitation of 1 mV was applied for the measurements.

The butterfly-shaped hysteresis curve in Fig. 2a manifests a type of negative tunnel junction magnetoresistance, i.e., $R_J$ for the antiparallel magnetization alignment between the FGT and CGT layers, $R_{AP}$, is lower than its parallel alignment counterpart, $R_P$. In this work we use the formula TMR= $\frac{R_{AP}-R_P}{R_P} \times 100\%$ to characterize the tunnel junction magnetoresistance, where the $R_P$ and $R_{AP}$ are taken as the $R_J$ values right before and after the magnetization reversal in FGT. The data in Fig. 3a yields a TMR ratio of −27%. Using Jullière's formula, TMR= $\frac{2P_1P_2}{1-P_1P_2}$, and taking the spin polarization in FGT to be $P_1 = 66\%$[39], one obtains $P_2 \approx -24\%$ at $T = 1.6$ K. Here, $P_2$ is the spin filtering efficiency of CGT, defined as $P_2 = \frac{I_\uparrow - I_\downarrow}{I_\uparrow + I_\downarrow} \times 100\%$, where $I_\uparrow$ ($I_\downarrow$) is the spin up (down) tunneling current.

As shown in Fig. 3b, the butterfly-shaped hysteresis becomes less pronounced when $T$ increases from 1.6 K to 50 K. This is followed by a vertical flip of the hysteresis loop when the temperature reaches 70 K. Fig. 3c displays the temperature dependence of the extracted TMR ratio. The magnitude of the negative TMR decreases with increasing temperature, and almost vanishes at $T = 60$ K, which is close to the ferromagnetic ordering temperature of a 2.1 nm thick CGT layer[38]. The TMR ratio becomes positive at 70 K and 80 K. It is also noteworthy that the TMR magnitude drops by a factor of about 3 as $T$ increases from 1.6 K to 20 K. Since the magnetization of the CGT layer should vary by less than 20% in this temperature range[38], such a large drop in TMR cannot be explained by the thermally induced demagnetization.

We notice that the coercive field of the FGT layer drops from $\mu_0H_c$=1.1 T to 0.7 T for the temperature interval of 1.6 to 20 K. As shown in Fig. 3a with a dotted line in the bottom two panels, if the magnetization reversal in FGT at $T$ =1.6 K took place at $\mu_0H$=0.7 T, rather than 1.1 T, the TMR ratio would become −11% (referred to as the hypothetical TMR ratio in Fig. 3c). This value is quite close to the −9% observed at 20 K. Similar hypothetical TMR ratios are also obtained for other temperatures. As depicted in Fig. 3c, they are not much different from the measured TMR ratios at $T \leq 40$ K. This suggest that the rapid drop in the TMR magnitude in this temperature range is not related to the $T$-dependence of the intrinsic conduction band spin-splitting. It rather can be attributed to a rapid decrease in the coercive filed of the FGT layer and a slow magnetization reversal process in the 2.1 nm thick CGT layer. In contrast, the magnetization reverses much faster in thicker CGT layers. For instance, the TMR magnitude of device B, in which the CGT tunnel barrier is 3.5 nm thick, only decreases by about 20% from $T = 1.6$ K to 20 K (See the supplementary material). In order to minimize the complications due to the large variations in $H_c$, here we introduce $m(T)$, defined as a ratio between the measured and hypothetical TMR ratios. It provides a more direct connection to the magnetization of tunnel barrier, or more precisely, the degree of spin splitting in the CGT conduction band. As depicted in Fig. 3d, the overall characteristic of the $m(T)$ data resembles the $T$-dependence of magnetization for a typical ferromagnet.

Fig. 4a shows that the TMR magnitude of the FGT/CGT/graphite junction drops very rapidly as the dc bias voltage increases. Similar phenomena have often been observed in other types of magnetic tunnel junctions, including the junctions with nonmagnetic tunnel barriers, such as Ge[40], AlO$_x$[41], MgO[42] and h-BN[39], and those with magnetic barriers, such as EuS[43], CoFe$_2$O$_4$[44], NiFe$_2$O$_4$[9]. Inelastic scatterings involving magnons[43,44] and defects[9,10] are generally believed to be responsible for the bias dependence. Energy dependence of the wavefunction decaying constant in the tunnel barrier, which is sensitive to the details of band structure, is also considered to be an important factor[9]. As shown in Fig. 4a, the TMR ratio becomes about −2% at bias $V=20$ mV. It is worth noting that at $|V| < 20$ mV there exist many peaks in the inelastic electron tunneling spectrum (Fig. 4b). These peaks can be attributed to low energy excitations, including magnons or phonons. Moreover, it is worth mentioning that the TMR ratios shown in Fig. 4a were measured with ac bias of 0.2 mV. The zero-dc-bias TMR reaches −30%, about 10% larger in magnitude than that measured with 1 mV ac excitation (see Fig. 3a). The sensitivity to the excitation level suggests the importance of low energy excitations in the spin polarized tunneling transport, consistent with a recent study of the magnon spectra in CGT[45]. As to the slight increase in the magnitude of TMR with increasing bias at $|V| > 60$ mV, it might be viewed as a precursor to the Fowler-Nordheim tunneling regime, in which the bias dependence opposite to that of the low-bias region was observed previously in EuS-based spin-filtering devices[46].

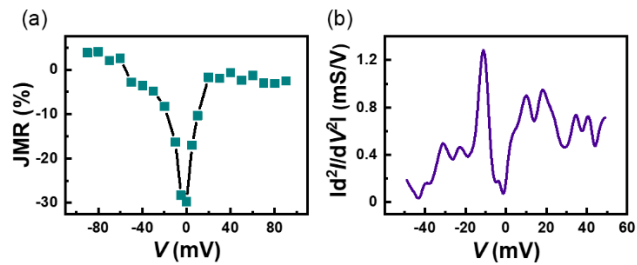

**Figure 4.** Bias dependences of the TMR and the inelastic electron tunneling spectrum of the spin filtering device at $T$=1.6 K. (a) TMR vs. dc bias voltage, measured with an ac current of 1 nA, corresponding to an ac voltage of ~0.2 mV at zero dc bias. (b) Bias dependence of $d^2I/dV^2$ in absolute value.

In addition to the inelastic scatterings due to phonons and magnons, it should also be noted that in a scanning tunneling microscopy study of CGT, Cr-Te antistites were identified to the main types of defects, resulting in a formation of in-gap electron states[24]. The low energy excitations of these defect states could serve as additional source for inelastic scatterings. Furthermore, as depicted in Fig. 2d, the zero-bias differential conductance $G_0$ approaches to a finite value (~5 μS) at the zero-$T$ limit, at which the two-magnon-assisted tunneling process is expected to be fully suppressed due to the vanishing magnon population. This indicates that the defect-related in-gap states could provide additional channels for the transport across the CGT layer. Since both CGT and FGT layers are magnetic, the defect-assisted tunneling process probably do not conserve the spins, and consequently the spin filtering efficiency and the TMR ratio are lowered substantially. This helps to explain that the extracted spin-splitting in CGT is much smaller than the theoretical values (see supplementary material).

The temperature dependence of TMR measured at zero-dc bias is worth further discussion. As shown Fig. 3c, the TMR ratio remains negative at $T < 60$ K, but becomes positive at higher temperatures (e.g., $T =$70 K and 80 K). Because FGT has a positive spin polarization[20,47], the negative TMR suggests that the tunneling current across the ferromagnetic CGT layer has a negative spin polarization if the carriers are injected from a nonmagnetic electrode. We therefore expect that the tunneling barrier for the minority spins is lower than that for the majority spins. This agrees with several first principles calculations that predicted the conduction band minimum in CGT belongs to the minority spins[23,26,27], but in contradiction to a few other reports[24,25]. As to the positive TMRs observed at 70 K and 80 K, it probably originates from the magnetic proximity effect between the CGT and FGT layers. Such temperatures are slightly higher than the ferromagnetic ordering temperature of the CGT layer (~ 60 K), but much lower than the $T_C$ of the FGT layer[28,29]. Evidence for interfacial proximity effects has been found in other CGT-based heterostructures, such as CGT/topological insulator[21] and CGT/Pt[48] heterostructures.

In summary, we have shown that spin polarized tunneling based on vdW heterostructures can provide unique insight into the electronic structure and magnetic properties of intrinsic magnetic semiconductor CGT. The detailed knowledge of the spin-dependent band structures of magnetic insulators is indispensable for a deep understanding of the experiment exploring their spin filtering effect and the proximity effects between magnetic insulators and a large variety of interesting materials, such as graphene[49], topological insulators[21], and superconductors[22]. Extending similar measurements to other materials should be helpful to the search for 2D magnetic insulators for realizing practical TMR devices and exotic quantum effects.

See on the online *supplementary material* (https://doi.org/10.1063/5.0102745) for the methods for crystal growth, device fabrication and transport measurement, and the additional experimental data and analysis, which includes transport properties of FGT, $T$-dependences of $R_J$ for devices A and B, estimation of tunneling barrier height, Fowler-Nordheim plots at different temperatures, estimation of the size of exchange splitting, inelastic tunneling spectroscopy measurement, TMR of device B, and $I$-$V$ characteristics of a heterojunction with non-magnetic electrodes.

**Acknowledgments**：This project was supported by the Strategic Priority Research Program of Chinese Academy of Sciences (Grant No. XDB28000000), the National Natural Science Foundation of China (Grant No. 11961141011), and the National Key Research and Development Program of China (Grant No. 2016YFA0300600). P.X. acknowledges financial support by National Science Foundation grant DMR-1905843.